# Toward Maximum Grip Process Modeling in Software Engineering

Sabah Al-Fedaghi
Computer Engineering Department
Kuwait University
Kuwait
sabah.alfedaghi@ku.edu.kw

*Abstract*—**Process modeling (PM) in software engineering involves a specific way of understanding the world. In this context, philosophical work is not merely intrinsically important; it can also stand up to some of the more established software engineering research metrics. The object-oriented methodology takes an object as the central concept of modeling. This paper follows from a series of papers that focus on the notion of *thinging* in the context of the analysis phase of software system modeling. We use an abstract machine named the Thinging Machine (TM) as the mechanism by which things reveal themselves. We introduce a more in-depth investigation of a grand TM that Signifies he "totality of entities" in the modelled system. We also present new notions, such as maximum grip, which refers to the level of granularity of the significance where optimum visibility of the model's meaning is given. The outcomes of this research indicate a positive improvement in the field of PM that may lead to enhance understanding of the object-oriented approach. TM also presents the possibility of developing a new method in PM.**

## I. INTRODUCTION

Process modeling (PM) refers to the modeling of a system's activity to capture the mechanisms involved in the workflow of a business process [1]. PM facilitates the development of systems in in several scientific fields, such as software engineering, enterprise modeling, knowledge modeling, simulation, and workflow systems. PM is "not done for one specific objective only, which partly explains the great diversity of approaches found in literature and practice" [2]. PM aims at capturing and representing, in a graphical model, the routine activities that comprise the processes of an organization to facilitate "a (sufficiently) complete and accurate snapshot of relevant organizational practices" [1].

In software engineering, business/organization PM is an important topic since it involves the following [3]:

1. Making sense of aspects of an enterprise and supporting communication among various stakeholders.
2. Computer-assisted analysis to gain knowledge about an enterprise by generating a simulation based on the model's contents.
3. Quality management through ensuring the work process's adherence to standards and regulations.
4. Model deployment and activation to integrate the model into an information system.
5. Using the model to contextualize a system development project.

In general, PM in software engineering grows increasingly more multidisciplinary, with various stakeholders dependent on each other for information. Failures in software system design produce results that are quite costly. Problems in communication and integration of subsystems have arisen from disparities in notional representations.

A key difficulty is the apparent weakness of PM framework. "Lack of consideration of engineering projects … has led to methodological challenges in creation of integrated tools and techniques for better analysis and management of complex projects" [4]. Current research in the area has concentrated on methodological approaches to functional specifications, flow descriptions, and system structure definitions [5], in which the specification is developed at different levels of granularity. Nevertheless, there seems to be at least a partial underlying theory for expressing the unified totality of processes. According to Riemer et al. [1], the philosophical underpinning that has already been applied to the field of knowledge management "has resulted in more nuanced teaching and practice".

Consequently, the need exists to "open up a space for future discourse and research into the nature, role and limitations of organizational modeling in general and process modeling in particular" [1]. Specifically, researchers should further explore a PM language that can be "expressive, formal enough, and easily understandable by experts and end users" [6]. The solution should incorporate the features of integration applied to multidomain activities and interactions (e.g., physical, informational, and technical activities). Additionally, the complexity of the description ought to be expressible through simple notations apply to diverse detail levels. Common understanding is facilitated through the use of one language that that can specify and communicate.

PM involves a specific way of understanding the world. "Modeling can take the form of abstraction, fictionalization, idealization, and also approximate representations of what is observable from nature" [7]. In this paper, we adopt diagrammatic PM. According to Frichot [8], a diagram is a thinking technology hat secures an understanding of more than the static relations between components and dynamic movements. Diagrammatic descriptions in software engineering can be expressive and easily comprehensible by end users. Users enrich descriptions of processes with diagrams, which allows them to easily identify inconsistencies, nonterminating conditions, and so forth [6].

Additionally, we adopt a model constructed on the philosophical foundations of Heidegger's notion of *thinging* [9]. Such a Structural underpinning may raise concerns. Bogost [10] argued against the position that "When speaking to the general public, terms like a *priori*, *Heidegger* and *anti-realism* ought to be avoided". According to the site ThingMesh [11],

> Martin Heidegger dealt with, at some length, the question of what is a thing? As the great philosopher noted, scientists (and their humble cousins, us engineers) are more concerned with practical, real-world ….problems and have little time for the philosophical paradoxes that such a simple question throws up.

Heidegger's philosophy gives an alternative analysis of "(1) eliciting knowledge of routine activities, (2) capturing knowledge from domain experts and (3) representing organizational reality in authentic ways" [1].

Nevertheless, in software engineering, PM involves complex processes that are often very difficult to capture. Many terminologies are often not well understood [12], and notations regularly contain duplicate features [13] that are loosely based on theoretical formalisms [14]. Accordingly, "Philosophical work is not only intrinsically important, but it can also stand up in terms of some of the more established research metrics to other types of IS [information systems] research" [15].

We embrace a way of enframing (i.e., gathering together; Heidegger's terminology) the world by treating it as a world of things through using an abstract tool called a thinging machine (TM) that views all components of a domain in terms of a single notion: TM. We focus on using PM in the business world to produce a visual description in terms of things that flow in a TM constructed of five simple "processes": creation, processing (changing), receiving, releasing, and transferring.

### *A. Related Works*

Process languages [16] include UML activity diagrams (ADs), business process modeling notation (BPMN), event-driven process chains (EPC) [17], role-activity diagrams (RADs), and the business process execution language for web services (BPEL-WS) [18]. UML ADs and BPMN currently allow the most expression of a process and are the easiest to integrate at the interchange and execution levels. The BPMN[19] "represents the high-level graphical representation of business processes easily understood by business analysts, and especially useful in communicating business requirements" [20]. Nevertheless, more work is needed to detail "how to best carry out the actual activity of modeling a process in practice" [1].

In general, most current approaches in the area of software engineering try to refine relationships among basic categories in domain ontology that involve determining the basic universals and relations. Specifically, object orientation has become the standard for the analysis and design phases of the software development process. Object-oriented (OO) PM offers various methods to create process diagrams.

Researchers claim that advantages of the OO model include simulating a designer's way of thinking [21] and "different kinds of object-oriented languages share the common feature of reducing complexity in the representation of technical systems and design processes" [22]. The success of the OO approach is correlated with a proliferation of OO technology. According to Duckham [23], "this proliferation has not always been complemented by a growth in OO theory. The surfeit of object-oriented analysis, design and programming techniques which exist are, therefore, necessarily highly subjective". According to Joque [24],

> Despite the obvious allusion to object-oriented programming in the naming of object-oriented ontology, there are few descriptions of the relationship between object-oriented programming and said ontology. This is especially unfortunate as the history and philosophy that surround object-oriented programming offer a nuanced understanding of objects, their ability to hide part of themselves from the world, their relations, and their representation in languages that in many ways challenge the claims offered by object-oriented ontology.

### *B. A New Approach*

The OO approach takes an *object* as the central concept of PM. By contrast, this paper is a sequel to a series of papers over 2018–2019 [25-57], which explore alternative PM based on the notion of a *thing* and *thinging*. We use an abstract machine named a TM and introduce more in-depth analysis of TM.

The benefit of this approach is that it may supplement the OO paradigm with new notations and facilitate further understanding of some aspects of the OO paradigm. There is also the possibility of developing a new approach to PM based on TM.

According to Heidegger [9], thinging expresses how a "thing things", which he explained as gathering or tying together its constituent parts. In this paper, we use philosophers' ideas (e.g., Heidegger and Merleau-Ponty) to achieve modeling with a process-based approach that emphasizes processes over objects. It focuses on dynamics, events, and flows rather than static objects.

We regard the world we build and the underlying world we live in as closely bound by the modeling relation [58]. Here, the model is structured in terms of its domain (a portion of reality), not the world in general. TM *carves the system up* to impose thinging as the structure ascribed to the modelled system. It is assumed that the world being modeled consists of existing thing processes that interact with each other. The model captures an abstracted representation of this world. [1].

There is an isomorphic (i.e., a shared effect) relation between the model and its corresponding portion of reality, except for added organizational fabric. If the model, for example, represents the potentiality of a business order that flows to the supplier, then, in reality, a business order that flows to the supplier is an actualized process. In this case, the model is, in the jargon of logic, "true". The model comes to life as a physical machine, in reality and is integrated into its original domain, also in reality (e.g., a software system replaces the manual process). The "physical (manual) process" now wears its developed model and the model comes to life as a thing that things (see Fig. 1). It is only in use that the modelled system begins to reveal its potentiality; its being is understood through its handling (i.e., if its flow is applied in reality).

The aim of TM modeling is to describe the composition of external things and events that interact. The processes are employed to facilitate systemized utilization of thinging engineering. The model serves as a tool for communication between developers and users, thereby helping analysts to understand a domain, providing input to the design process, and documenting purposes.

## II. Thinging Machine (TM)

We view the world as a world of things, including data, materials, events, money, feelings, and so on (see Fig. 2). Paraphrasing Bryant [59], corporations such as the Coca-Cola Company, for example, are no less things than quarks, acorns, or stars in the way that electrons are no less things than rocks. All are being created, processed, and transported and all create, process, and transport, as will be described next.

Things, here, refer to beings that flowing within a TM where TM acts as a subject and the thing acts as an object. It handles things (see Fig. 3), the stages of which can be described as follows:

**Arrive**: A thing reaches a new machine.
**Accepted**: A thing is permitted to enter the machine.
If arriving things are always accepted, Arrive and Accept can be combined as a **Received** stage. For the purpose of simplification, the examples in this paper assume the received stage exists without loss of generality.
**Processed** (changed): A thing undergoes some kind of transformation that changes it without creating a new thing.
**Released**: A thing is marked as ready to be transferred outside the machine.
**Transferred**: A thing is transported somewhere from/to outside the machine.
**Created**: A new thing is born (created) in a machine. The term Create comes from creativity with respect to a system, e.g., constructed things from already created things, or emergent things appear from somewhere.

The following discussion familiarizes, first, the part of the TM model that emphasizes things, delaying until later discussion of the thesis that processes are things and things are processes. Note that the word *process* is meant in the generally used sense (machine, procedure, task, undertaking, etc.), and not as a stage in a TM that denotes change in a thing. In general, we orient ourselves to what Bryant [59] calls "distributed agency". A thing assumes the role of an agent when it is a machine. Using Bryant's [59] ideas, things can be simultaneously viewed as unities and as machines. In both cases, they are assemblages composed of other things and machines. Machines draw on other machines to produce themselves.

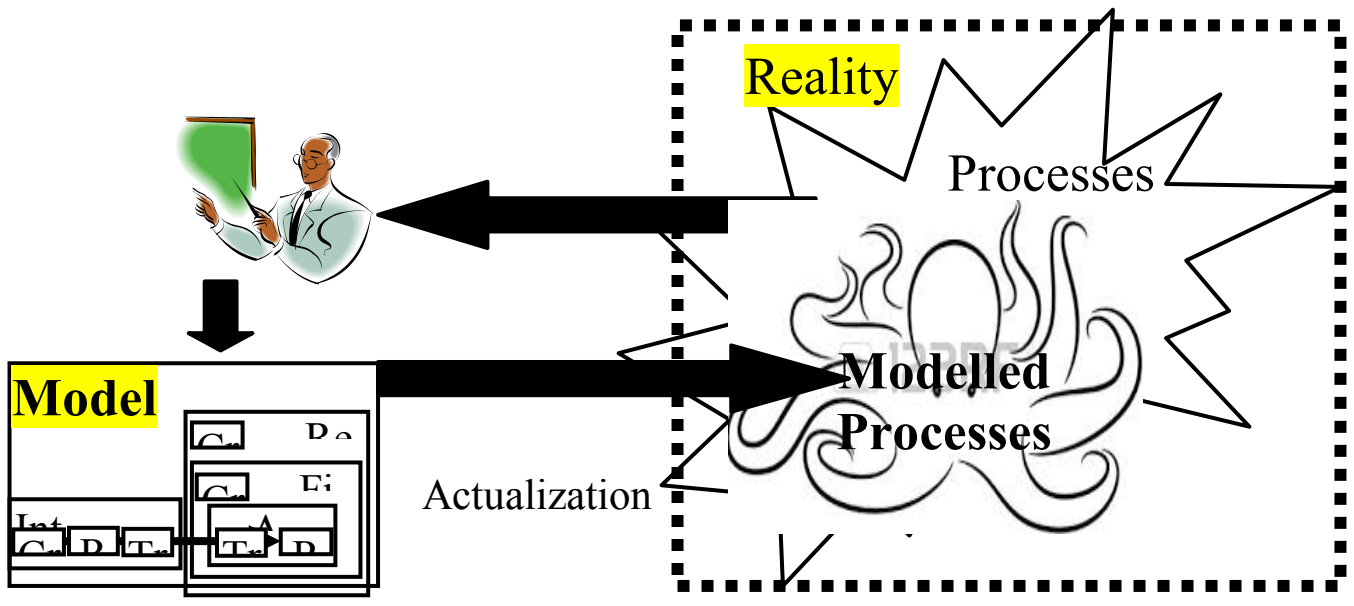

Fig. 1. The manual system "wears" the produced model.

### A. Things

A thing is defined as that which can be created, processed, released, transferred, and received. It encounters us through its givenness (Heidegger's term). "In contrast to object-orientation which represents things as quantifiable objects to be controlled and dominated Heidegger's definition of a thing encompasses a particular concrete existence along with its interconnectedness to the world" [60]. Accordingly, the TM model is attuned to the unique ways in which entities present themselves or exist.

Operations on things are limited to creating, releasing, transferring, receiving, and processing. An operation (stage of flow) denotes some type of "cooking" of things. It is also not stage-referential (e.g., create [process]). The stages manifest by means of the flow of things. The flow ensures continuity (unity) of visualization of the processes specified in the narratives of the system.

This thing orientation can be applied to business processes/things. We let the thing assert its "existence," along with its interconnectedness to the world. Instead of segmenting things into classes, packages, components, states, etc., the software engineer–thing relationship is analogous to an artist who storyboards the thing as it plays its part in the business theater. The resultant drawing then acts as the script in specifying the thing's flow (in the play).

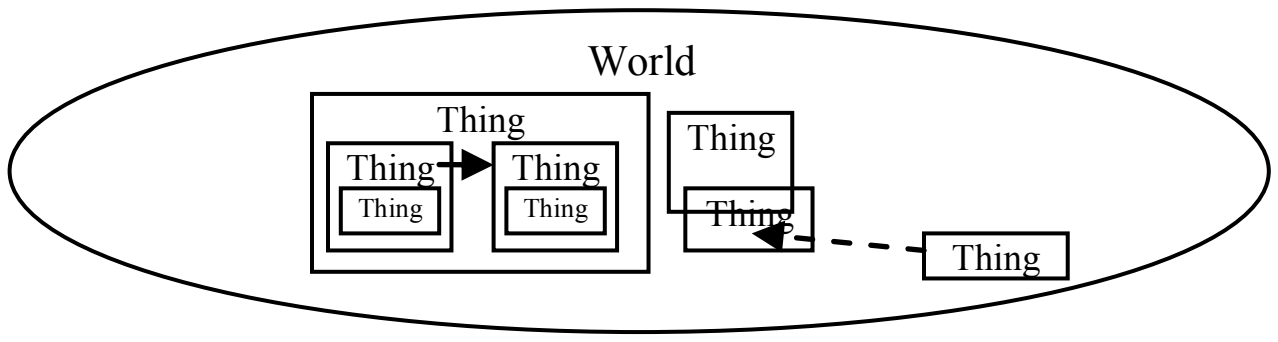

Fig. 2. The world of things.

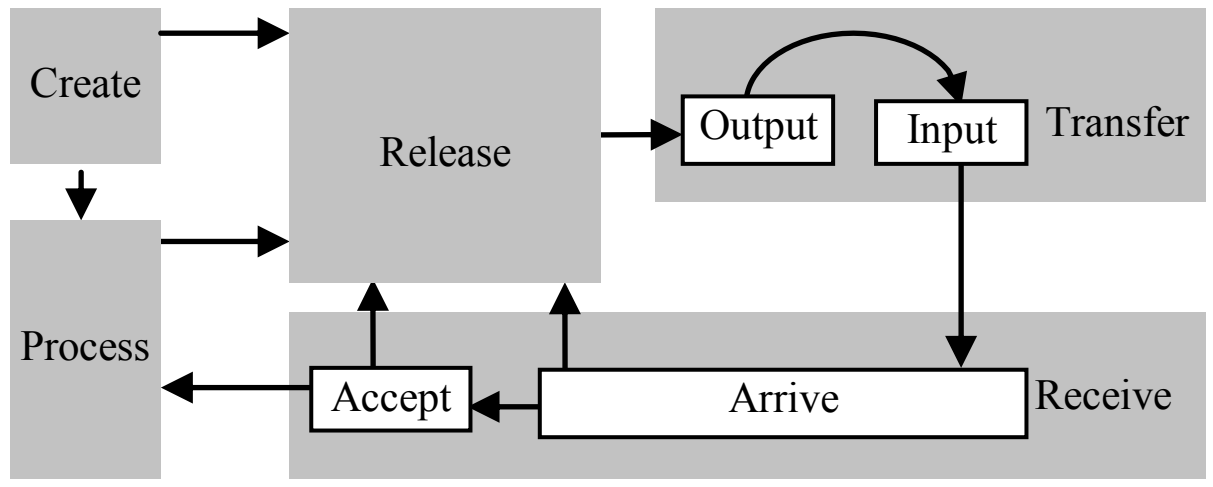

Fig. 3. Thinging machine.

### B. *Flow of Things*

There are two types of “movements” in Fig. 3:

(i) Flow (solid arrow): Flow is a spatiotemporal event, but at the start, we are interested in identifying trajectories of flow in a machine in rest. The flow of a thing signifies its *conceptual* movement, from one machine to another or among stages of a machine. Conceptual flow is not necessarily physical flow. For example, in a manufacturing assembly line, if a device (thing) arrives at a position where two robots process two parts of the device, then there are simultaneous flows (transfer, receive, and process) to the two robot machines (conceptual space).

The difference between the flow of abstract things and physical things is, quite simply, that the abstract things are “invisible” (Merleau-Ponty’s term [61]) and cannot flow directly to the world of visible machines. The flow movement of things actually produces an “existential space” (Merleau-Ponty’s term).

(ii) Triggering (dashed arrow [will be shown later]): Triggering (denoted by a dashed arrow) is a special type of flow. A flow is triggered if it is created or activated by another flow (e.g., a flow of electricity triggers a flow of heat) or activated by another point in the flow. Triggering can also be used to initiate events, such as starting up a machine.

(iii)

**Example**: A TM offers a different way of conceptualizing a modeling approach that rejects the subject/object structure. It is a way of thinking that guides the technical understanding (care of handling) of inspired practice. It is a way to lessen (Heidegger’s terminology) the habituated handling of software engineering work and to rethink the affairs that arise during modeling.

Consider how the OO model describes a *person*, as shown in Fig. 4. It is a class with attributes and methods. The behavior is mean to be described in another diagram (activity diagram) with completely different types of notations. We will not represent these different types for the class *person* because they are familiar to software engineers.

Fig. 5 shows the corresponding TM model. A person is four (sub)TMs:

(i) The person him/herself TM (circle 1),
(ii) Work TM (2),
(iii) Eat TM (3), and
(iv) Name TM (4).

The whole diagram is a machine called the grand machine (GM).

Additionally, the behavior of this machine is modeled in terms of events. Events are things that can be created, processed, received, released, and transferred. As an example, Fig. 6 shows the event *A person goes to work,* which is a machine that includes

(i) The submachine of location of the event (subdiagram of Fig. 5),
(ii) The time submachine, and
(iii) The event itself submachines.

For simplicity’s sake, in the examples deployed in this paper, events are represented only by their regions.

An event extends across time, from a beginning to an end. Its time ontology embraces *things* and their *flows*. Events are unified when the diagram depicts flows within the stretch of time of the events, from creation until they run their course in time. When actualized, the diagram is the “content” of an event; that is, the diagram is *to be executed* and the event *to execute*.

An event is also a flow in the spatial location specified in the static model over time. A meaningful event refers to an event that manifests itself only as a step in the entire behavior of the modeled system. This behavior is composed of distinct things—some stationary, others moving—in various flow streams at different machines. At any given moment in time, the collection of events occurring and things’ positions form an instance of the system state. The states can form a three-dimensional representation of the GM that represents the “history” of the system.

Fig. 7 shows four selected events in the person machine.

A person appears in the system.
A person goes to work.
A person eats.
A person’s name is given.

Fig. 8 shows the chronology of these events.

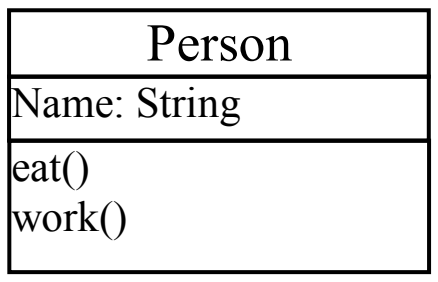

Fig. 4. The class Person (adapted from [62])

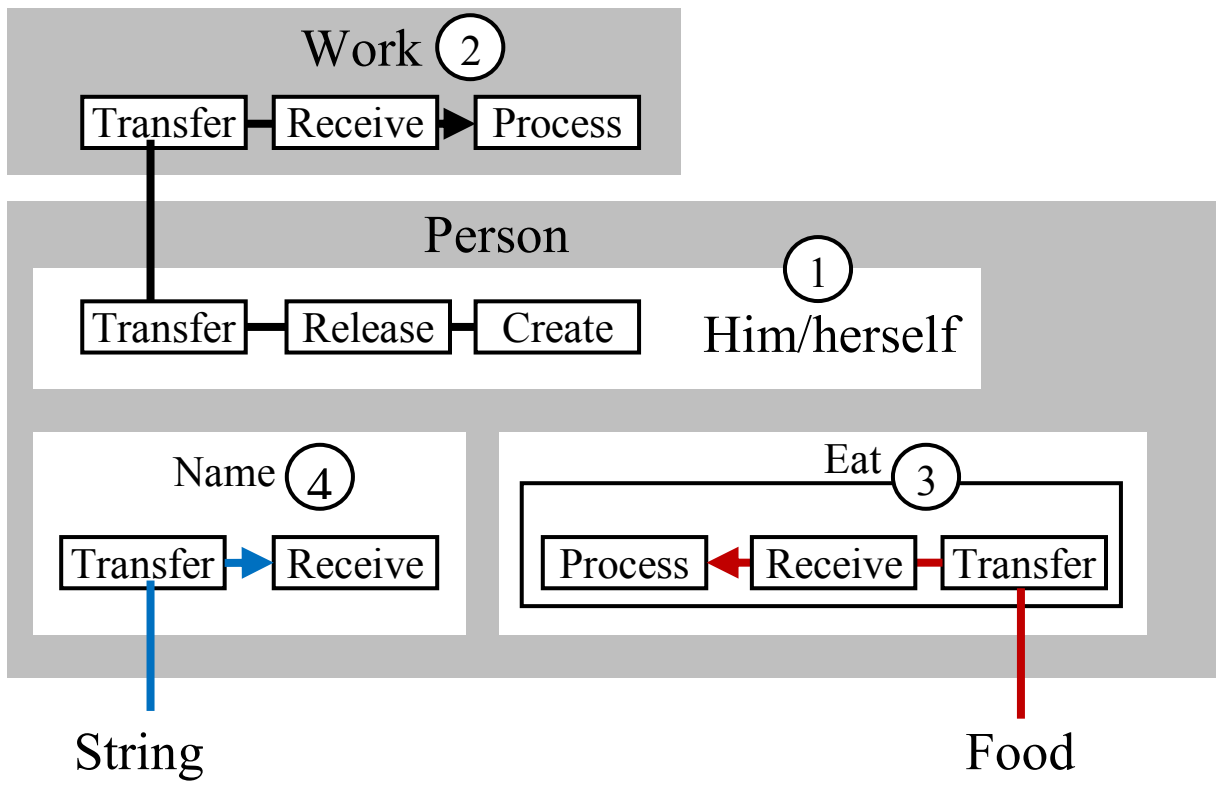

Fig. 5. The TM description of the example.

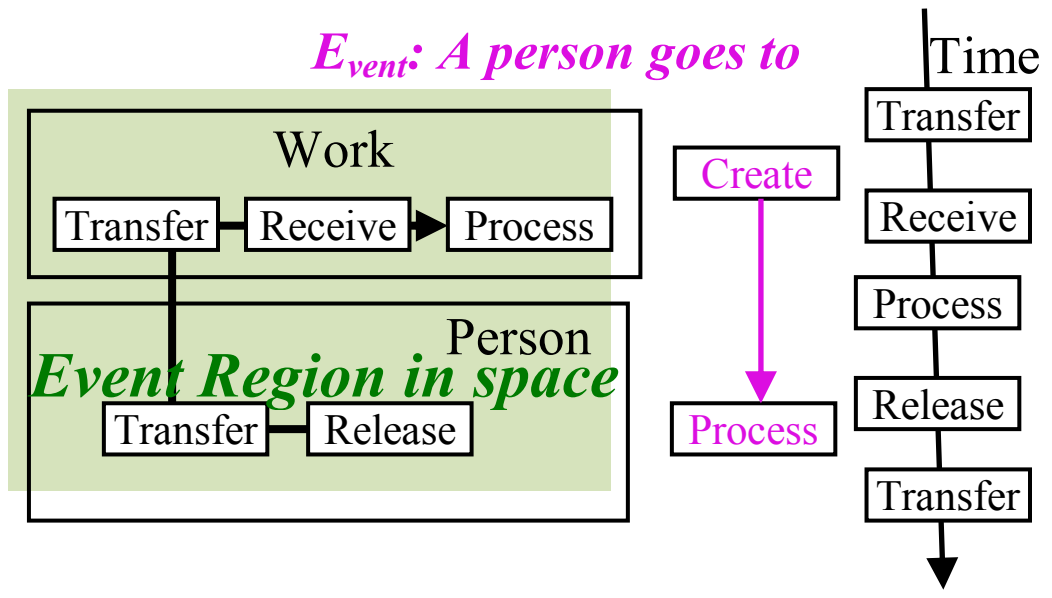

Fig. 6. The event *A person goes to work.*

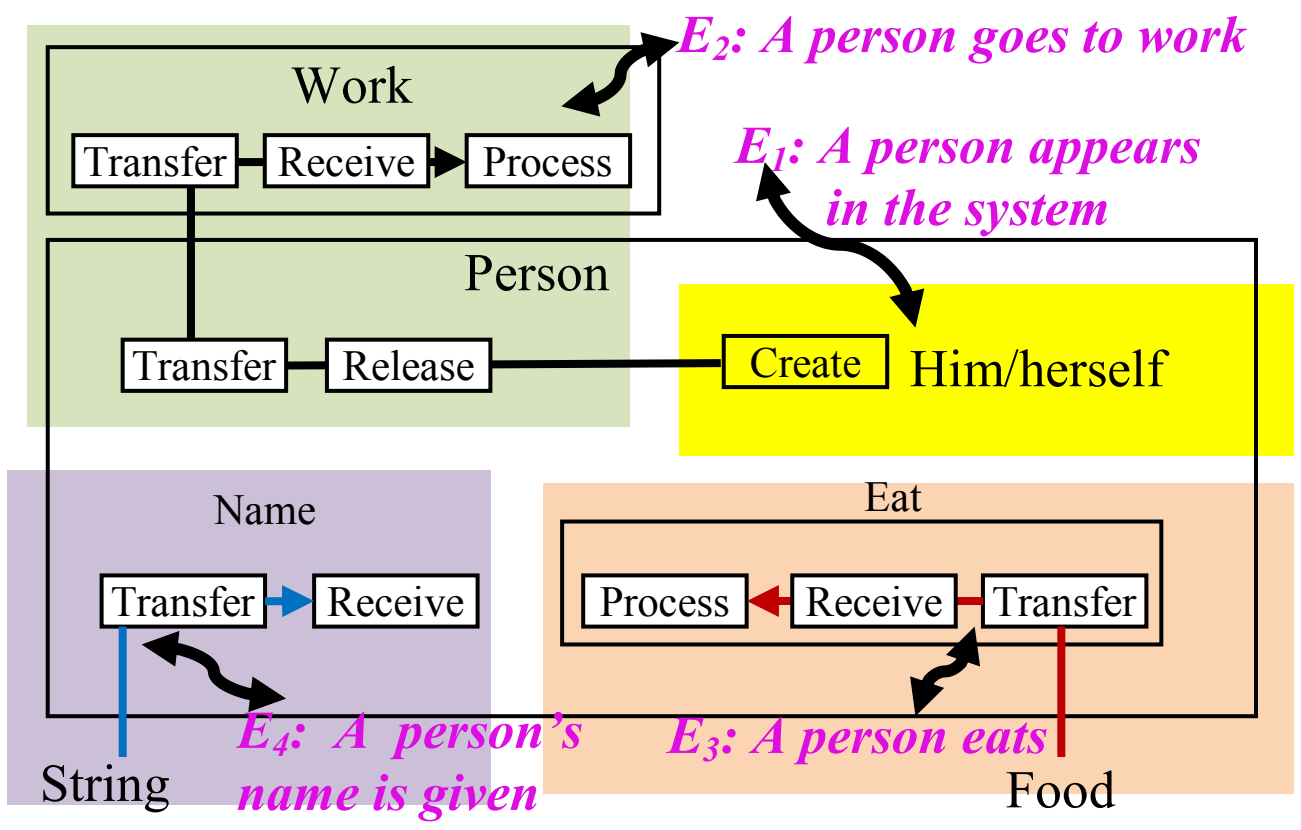

Fig. 7. The TM description of the example.

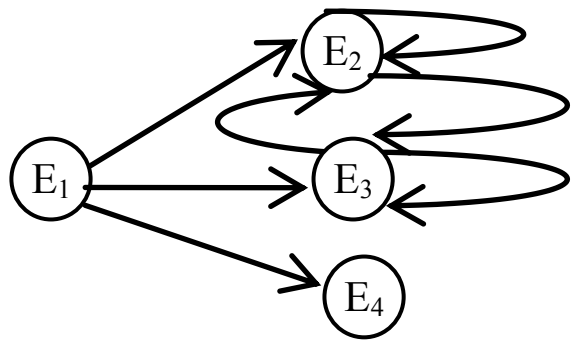

Fig. 8. Chronology of events.

Now, the reader is asked to contrast the series of TM diagrams with how the examples would be expressed in UML diagrams, including behavior diagrams. Observe the uniform development of the static and dynamic representation of the system in TM, and that it does not get lost in a sea of notions (e.g., activation boxes, activity, actor symbols, package symbols, lifeline symbols, synchronous message symbols, message symbols, and reply message symbols).

In TM, a person is created (i.e., an object is initialized) by the appearance of a person in the system. The creation of a person causes the creation of name and the fact of eating that may initially be nulls. Conceptually, work is not an attribute of a person; it is another thing that the person flows to. The description compels the introduction of two new things (i.e., food and string). It is not difficult to imagine a technical development of a language that,

(i) creates an instance of a person with a name and eating activity analogous to an object with attributes and methods, and,

(ii) creates a work function that accepts a person, which may have its own attributes. Such technicalities are based on clear conceptualization.

## III. Things Are Machines and Machines Are Things

According to Bogost [63], a machine is something that operates. It is defined in terms of its organization. In a TM, a machine operates by creating, processing, receiving, releasing, and transferring things. *A tree* is a machine "through which flows of sunlight, water, carbon dioxide, minerals in the soil, etc., flow. Through a series of operations, the machine transforms those flows of matter, those other machines that pass through it, into various sorts of cells" [64].

In the TM approach, a thing is not just an entity, but it is also a machine that handles other things. Heidegger's [65] modes of being mark two irreducible aspects of every object:

- **Present-at-hand** refers to conceiving *being* in terms of tangible "objects of some sort of discussion or perception, and is recognized by a specific shape and color and texture" [66].
- **Ready-to-hand** is the mode of beings in themselves: "an invisible thing radiating its being is dissolved into the world" [66].

This ready-to-hand mode can be interpreted as a thing that moves and changes or breaks relationships with other things [64]. "The being of an object lies not in whatever qualities it happens to manifest or actualize, but rather in what an object is capable of *doing*; its effects" [64; italics added]. Bryant [64] declares, "All objects can be understood as machines".

Machines are also "cooked" (i.e., operated on); that is, created, processed, released, transferred, and received. The intertwining with the world is accomplished through both of these modes of being, as an entity that flows through machines and as a machine that other things flow through. The unity of things/machines inhabiting the system does not arise from applying traditional categorization, properties, and behavior, but through the activities (create, process, etc.) and flows of the inside things.

Things are machines and vice versa. A TM is a type of process (in the general sense of process), where the process is "fixed" as creating, processing, releasing, transferring, and/or receiving. This is Heidegger's idea of being as a process (specified as "of emergence/becoming"). For Heidegger, "nature is seen as first and foremost 'productivity' or process, and only derivatively as 'products' or things" [67]. From the classical perspective, the thesis of thing/machine may be connected to the notion of materials and form, however, in a TM they are two sides of the same coin. In a TM, this can be viewed as things that are transformed into machines and vice versa, reminding us of Escher's works such as *Angels & Demons*.

The *jug* is not a lifeless object; rather, it presents itself as a process. It gathers up, holds, offers, and pours drinks. Each of these is a subprocess in the thing/process *jug*. In object orientation modeling, this picture is wrapped up in a vague connection of class/object, properties, and methods described in terms of multiple diagrams with many kinds of icons and arrows. The idea of the thing/process is not far from the OO model, but it does not reach its maturity by merging the notion of object and process.

Accordingly, an alternative process-based way to view the world is a world of machines where things in Fig 2 are replaced by machines. For example, the OO model (Fig. 4) is regarded as describing *personingness* machine (process in the general sense) that interacts with naming, eating and working processes. Such a process-based (object-free) description will be left for future research. In this case, the Heideggerian' notion of inert Being finds an alternative conceptualization as dynamic process-infested flux. In this composition, everywhere there is process and there is a processing machine.

In general the thesis, *Things are machines and machines are things* gives us a tool to handle things by considering things as processes. A thing is a process implies that the TM model is a "blind" process that functions through sensing itself in reality. For example, if an order in placed by the customer then this is sensed by the machine, checks that the order is OK, then "anticipate" its arrival to the supplier, and if the order does not arrive the machine takes an action, etc. The TM model works through its senses (sensors) as a human being. It is an agent reacting to situations.

## IV. The Grand Machine

The entire TM diagram of a system is a machine called the GM. The GM is a "virtual system" in the sense of Merleau-Ponty's "virtual body" [68], where we can explore, say, our hands in space before actually moving them [61]. Similarly, the GM is an exploration of the "body" of the system before injecting movement into it as we superimpose its dynamism through events.

The static GM is where things are projected into "space". This space differentiates things and their *potential* flows. Things flow in the GM as abstract "river basins", which form the great "territories" inhabited by machines.

Time is incorporated in TM to "dress" the static model with a succession of events and simultaneity. In our incorporation of time into the static diagram, the stages of the machine (elementary events) may be grouped into events more easily mapped to meanings. The resultant GM system appears as a grouping of "multiperceptual" (inputs/outputs) machines with spatiotemporal patterns.

The GM is the whole of what is there, of what happens. The static model (without events) may include contradictions, as in a one-lane street that permits flows in both directions. The dynamic description solves that contradiction (e.g., flow in one direction from time 00.00 until 11:59 and flow in the opposite direction from 12:00 until 23:59). Hence the behavior of the system involves time, space through the event of *change itself*, and other properties such as intensity, suddenness, etc. The event indicates the presence as a now (time) and a here (space), as shown in Fig. 9.

The GM, including its events and event chronology, acts as a functional focus of the system activity, a place where all things flow and are processed to further create actions. It provides the representational nexus for controls the operations of the system and interfaces with the rest of the actual processes in reality. Each submachine performs its own task, taking care of a certain part of the flow (i.e., it performs the creation, processing, and transferring of certain things), oblivious to actions in other submachines. There are starting machines and monitoring/feedback machines, etc., but no submachine holds a plan or view of the system as a whole like the GM. Some submachines collectively use things (e.g., data) passed on from the other submachines to form an accumulation of submachines. The totality of the GM is a self-monitoring system that processes itself and acts in accordance with a feedback cycle as illustrated in Fig. 10.

**Example**: According to Gómez and Sanz [69], Heidegger differentiates two kinds of being, the readiness-to-hand and unreadiness-to-hand. A hammer has two different modes of being: a hammer hammering a nail or a hammer in a drawer. In this picture, there is no need for representation, but rather there is continuous sensory-motor immersion in its reality. The agent captures reality in the form of patterns (see Fig. 11).

Figs. 12-14 show the corresponding GM, its behavior in terms of selected events, and the diagram of the chronology of events. The TM modeling can be viewed in the context of the embodied cognition theories (Merleau-Ponty), as shown in Fig. 15. Perception becomes the means through which consciousness establishes itself as an integral part of the world.

This perception is not a channel that simply filters in information from a separate environment, but it is, rather, a kind of interconnectedness that allows for a simultaneity, in which one perceives the world through observation and interaction and experiences the world revealing itself through its very essence (see Fig. 16) [70].

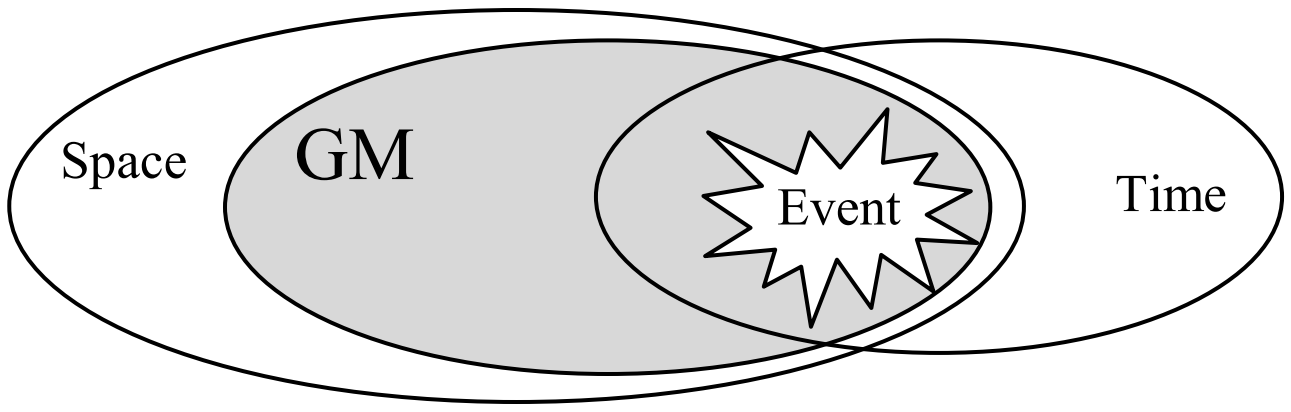

Fig. 9. GM is based in space and events are super imposed on it to incorporate time.

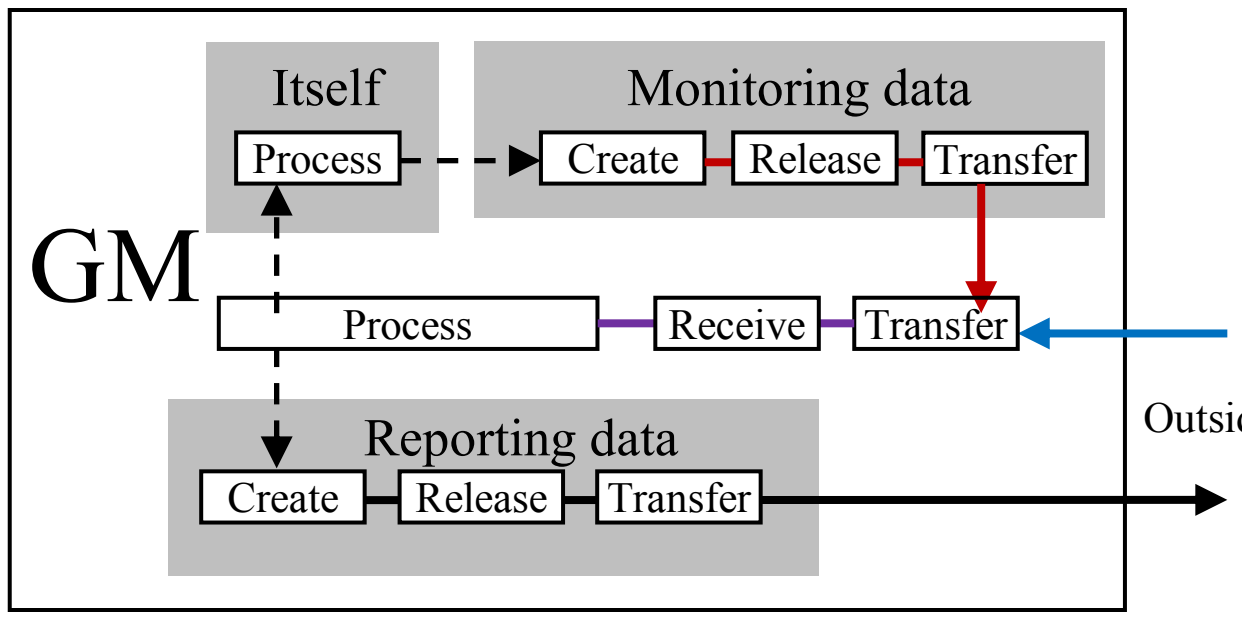

Fig. 10. The GM as a self-monitoring system.

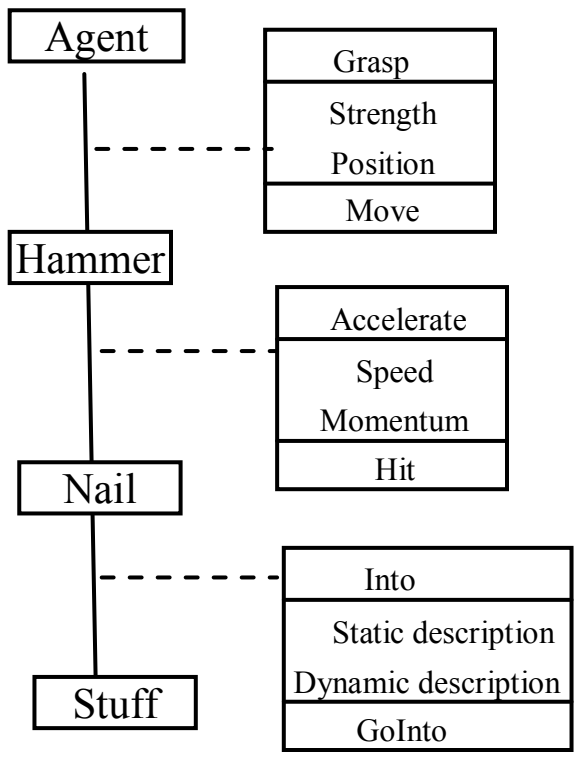

Fig. 11. The hammer, the nail, and the stuff (adapted from [69])

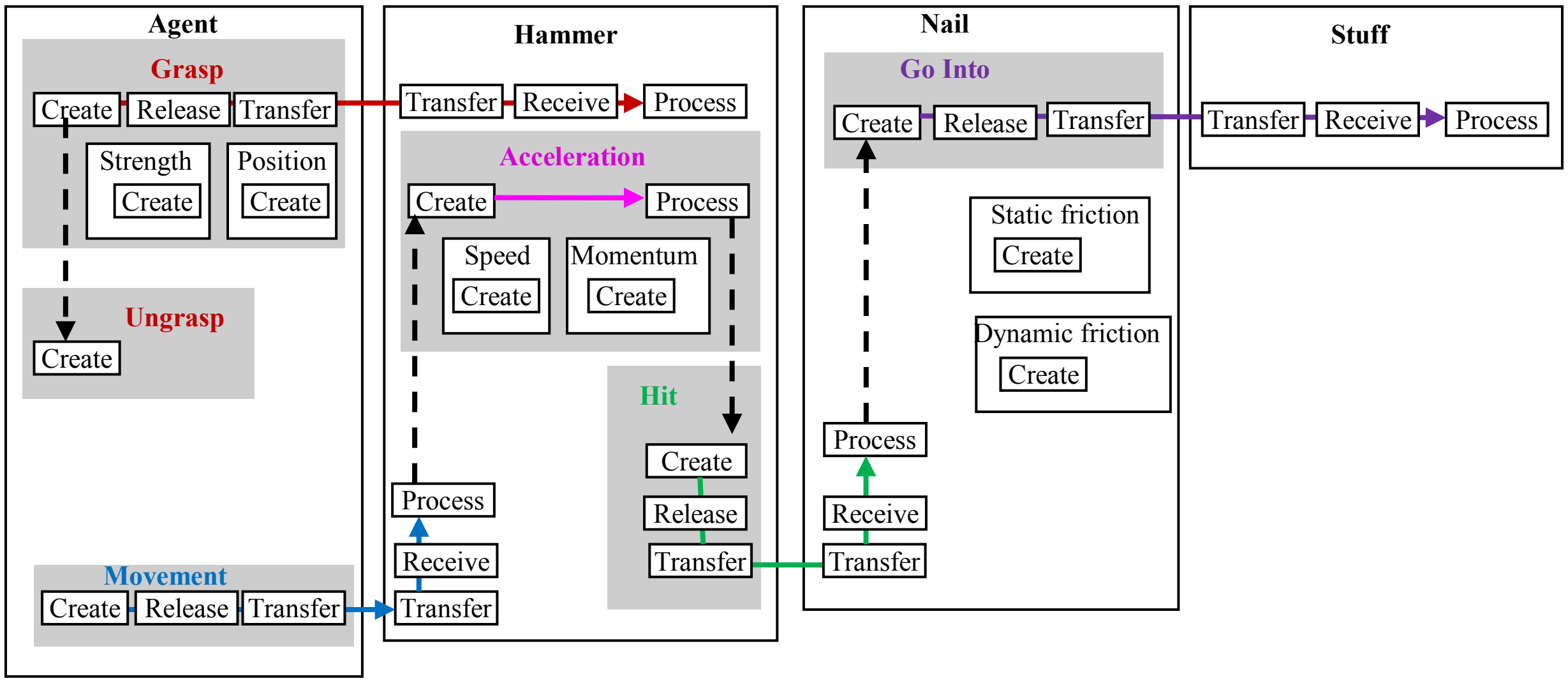

Fig. 12. The TM model of the hammer, the nail, and the stuff.

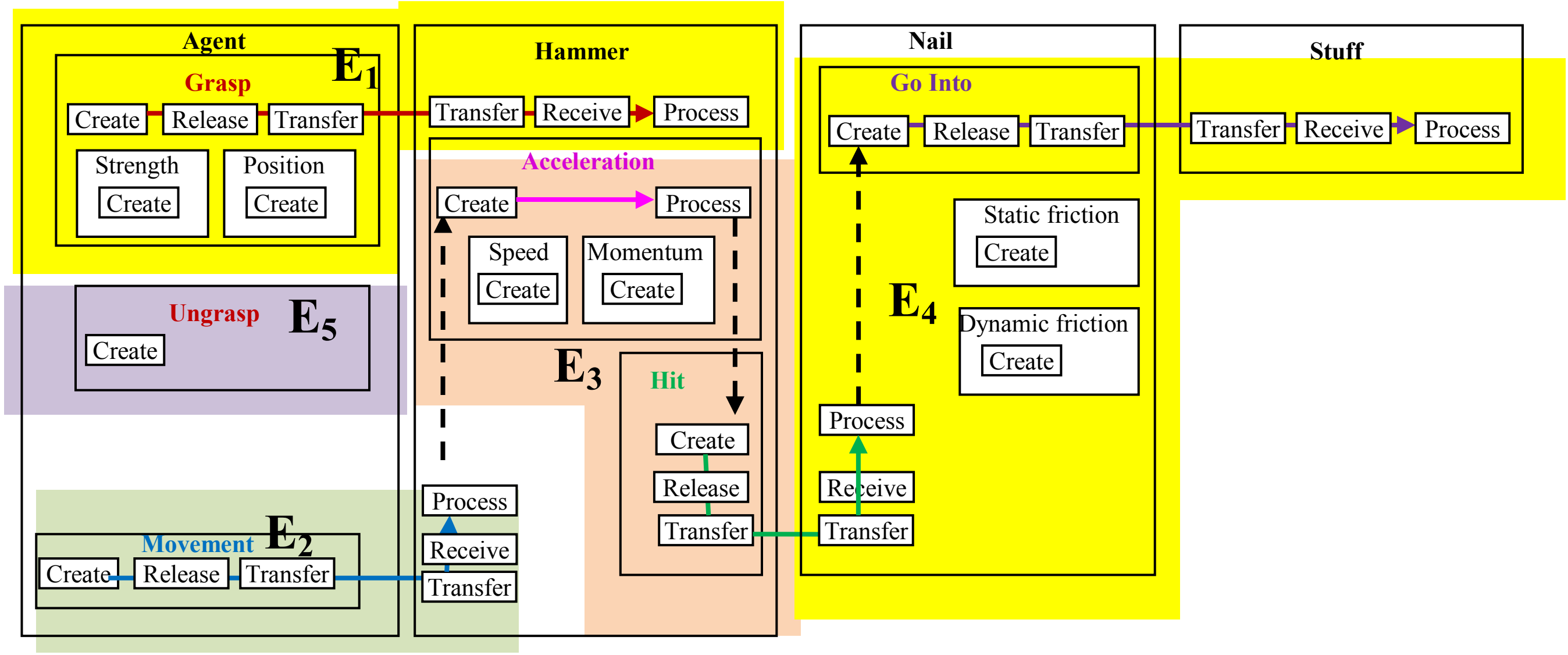

Fig. 13. The events in a TM model of the hammer, the nail, and the stuff.

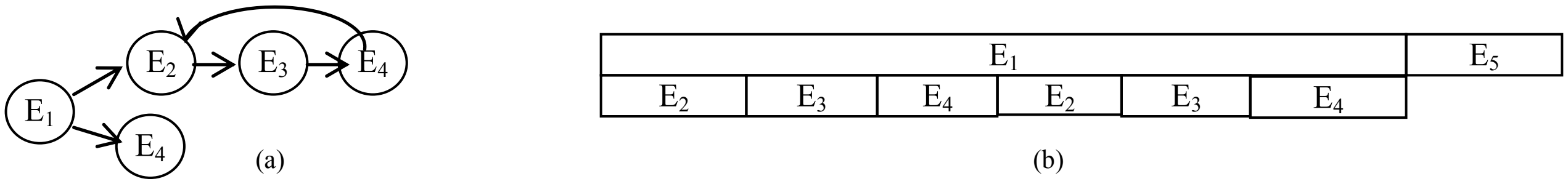

Fig. 14. (a) The chronology of events in the TM model of the hammering; (b) Event timing.

The behavior of the GM (system) is constructed from changes in submachines capable of functioning with relative autonomy while maintaining the same (static) relationship among themselves. The GM (including all diagrams) works as follows:

- The grasping machine works until the opening of the hand (ungrasp machine).
- While the grasping machine works, the movement machine operates.

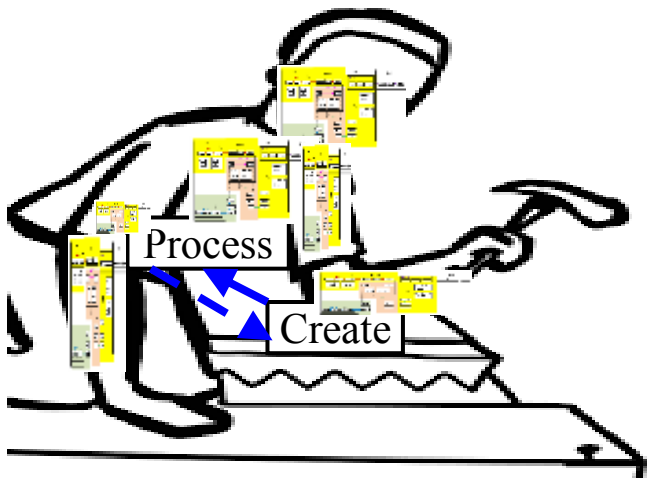

Fig. 15. Representation of hammering in the embodied cognition theory.

- The hammer machine is activated by input from the grasping and movement machines and experiences the acceleration machine to generate hits.
- The hit stimulates the nail machine to drive the nail in.
- The stuff machine receives the nail.

Each machine does its job. The grasping machine does the grasping, regardless of whether movement is generated. The movement machine does not care whether the nail goes in must include the hammering person, as a controlling agent. As part of the GM, the person may feel that the nail does not go in (i.e., the GM gives this monitoring data). The data are processed and appropriate action is taken (e.g., firmer grasp and faster movement).

## V. Thing/Machine Modeling

A software engineer's aim is to capture a description of a process. Thus the engineer has grapples with various philosophical issues, such as skepticism about appearance and reality; how does he/she know that what appears is real and so forth? The engineer distinguishes erroneous/veridical appearances by applying the description to the modeled system. Nevertheless, he/she starts from a point similar to that of philosophers, as in the case of searching for ontologies and theories of objects and their ties, to provide criteria that aids in distinguishing different types of objects (concrete and abstract, independent and dependent) and their connections (relations, dependencies and predication).

The starting point consists of vague concepts of how a process may be constructed from a coherent deformation of available significations (Merleau-Ponty) that are the individuations of machines. Such ideas, in the phenomenological approach within computer science, have been endorsed in the works of numerous writers in the research literature (see [71]).

### A. *Intentionality and Modeling*

Hence, the starting point of modeling involves *intentionality*, which denotes the idea that every mental phenomenon includes something as an object within itself (e.g., in presentation, something is presented; in judgment, something is affirmed or denied; in desire, something is desired and so on). In a TM, we can say that every perception phenomenon includes creating, processing, releasing, transferring, and receiving things.

Accordingly, the question "what is modeling *about*?" ought to be rephrased as "*what things are created, processed, etc.*?" in time and space in a model, as, for example (see Fig. 16), when directing one's sight at a physical box. Fig. 17 shows the involved events, where the two sides appear (Events 1 and 2) and meet (Events 3 and 4) to realize the act of perceiving (Event 5).

This thing-/machine-based modeling emphasizes the "active" nature of modeling, where we address the full richness of things pursuant to Merleau-Ponty's active nature of perception. Merleau-Ponty emphasized the dominant role of the body perception in engaging with the world. The TM model applies this view, to some degree, in modeling.

The following example illustrates how intentionality plays its role in a thing/machine process.

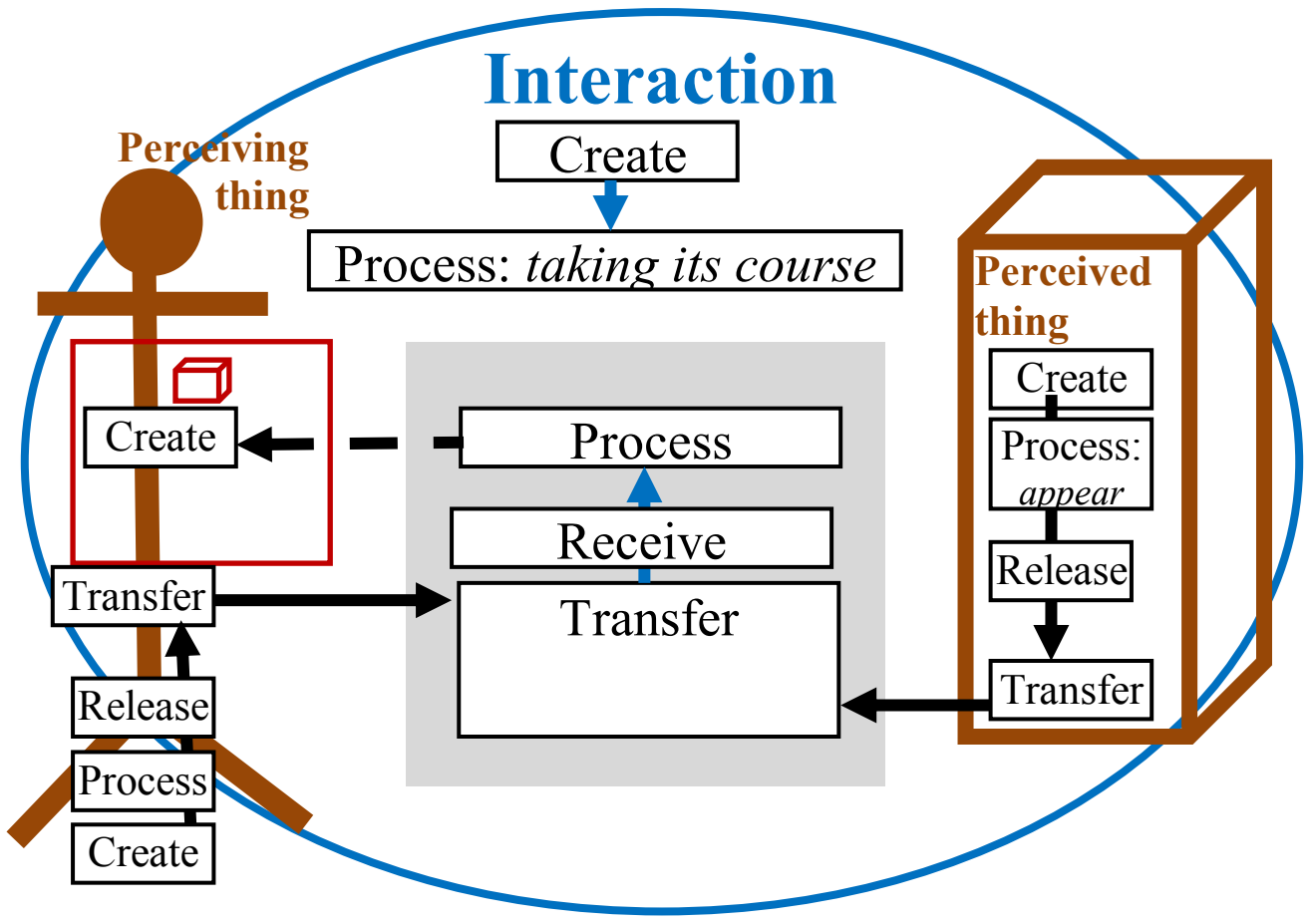

Fig. 16. The thing box is created, processed, received, released, and/or transferred.

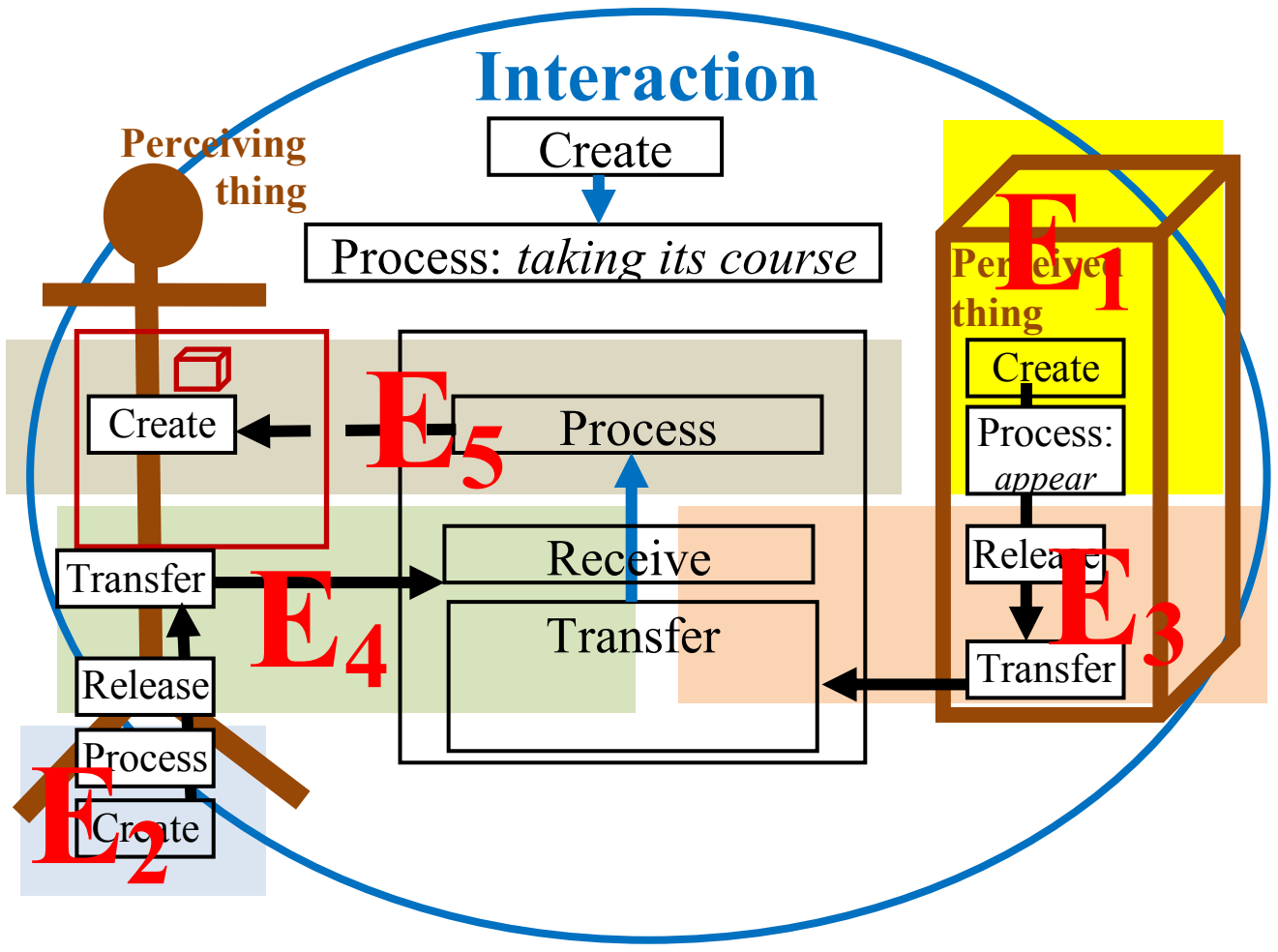

Fig. 17. The dynamic picture of perceiving

**Example:** Heidegger takes the example of the communion chalice, where silver is the material shaped into the form of "chaliceness". According to Heidegger,

> Silver is that out of which the silver chalice is made… The chalice is indebted to, that is, owes thanks to, the silver out of which it consists... The sacrificial vessel is at the same time indebted to … idea of chaliceness… a third that is above all responsible for the sacrificial vessel. It is that which in advance confines the chalice within the realm of consecration and bestowal…. Finally there is a fourth participant in the responsibility for the finished sacrificial vessel's lying before us ready for use, i.e., the silversmith. [72].

According to Heidegger, instead of seeing the silversmith as the agent that "effects" the production of the chalice, the model would view the careful consideration of the silversmith, to reveal the chalice's coming into becoming. The abstraction of "chaliceness", and the context in which the chalice will serve, is the method that allows the chalice to come into being [73].

Fig. 18 models such a process, where the silversmith's intentionality triggers the idea and thereby reveals the chalice's coming into being. The revealing starts with finding the silver and the form of the chaliceness and providing the melting/forming that give birth to a silver chalice that becomes a sacrificial vessel.

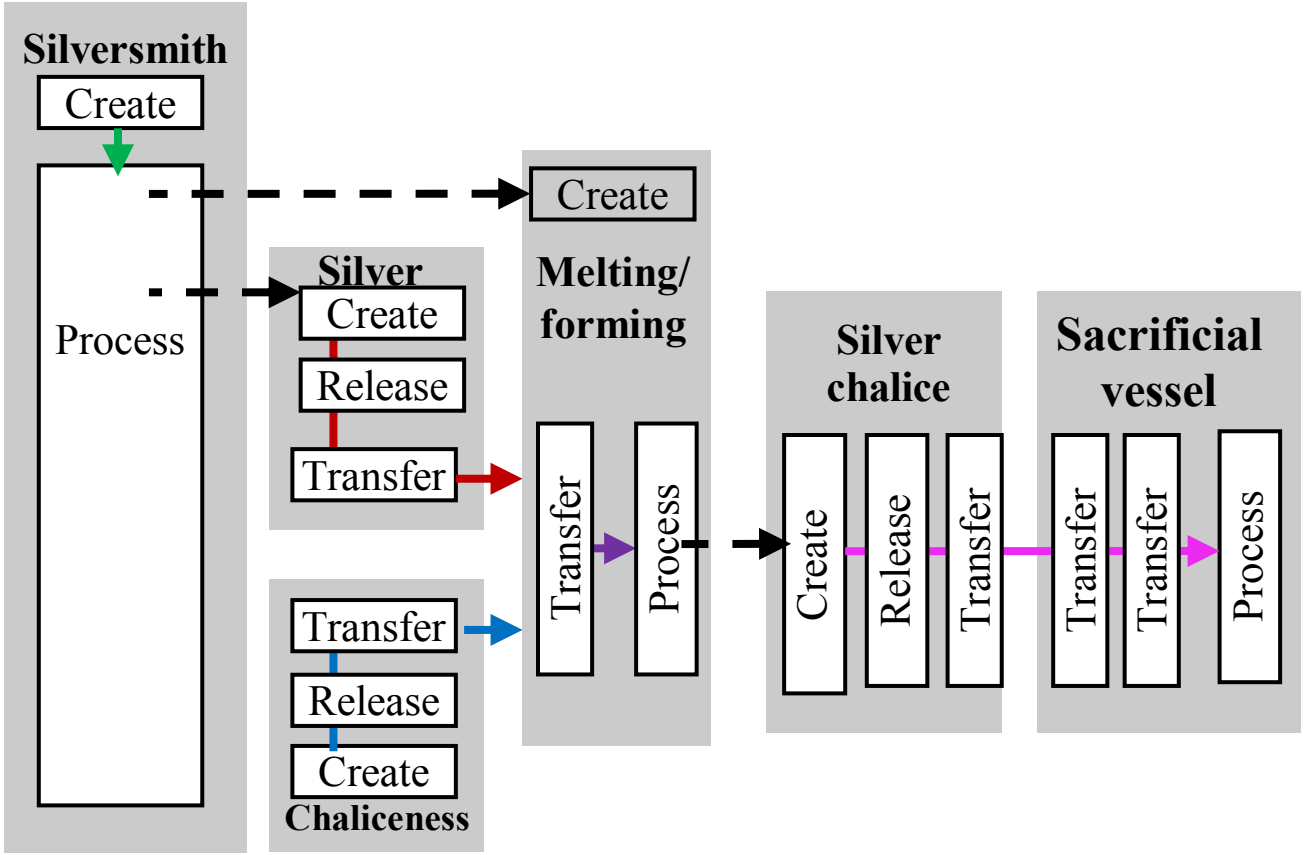

Fig. 18. TM description of Heidegger's example of the communion chalice.

### B. *Maximum Grip in Modeling*

According to Merleau-Ponty [68], for each thing in the world, "as for each picture in an art gallery", there is an "optimum distance" from which it is best observed. In terms of the notion of maximum grip, the observer places him/herself relative to a phenomenon to achieve optimal access. If he/she moves too close, there are too many details; if he/she moves too far away, the details are lost.

The lived body maintains a way of finding a better grip on the world [61]. This can be projected as a model's (hence, the modeler's) way of finding a firmer grip on the world's processes. One aspect of such a grip is the level of granularity of the significance where maximum visibility of the model's meaning is optimally given. Maximum grip indicates the inclination of the model, when utilized, to approach best results at the gestalt (overall) level.

Rietveld and Brouwers [74] discuss the phenomenon of the tendency towards an optimal grip on a situation in real-life situations in the field of architecture. An optimal grip concept is helpful for interpreting descriptions of complex situations in the design process. They distinguish between (a) a grip on visual perception, (b) a grip on the design, and (c) a grip on "how to design". According to the authors, the optimal grip concept is helpful for interpreting descriptions of complex situations in the architectural design process. The designer constantly zooms back and forth between various scales and aspects of different things to integrate and grasp the totality of the situation [75].

In the OO approach, to capture, say, a scene, the designer orients the scene backward and he/she turns his/her head to capture whether there is an object in the scene. The purpose of turning his/her head back/front is to grasp it piece (object) by piece and to avoid being dazed or disturbed by the wholeness of the scene. The world consists of recognizable objects with properties (e.g., size and color), and each must be registered separately. Accordingly, the designer turns back to capture the properties and behavior, as well as the relationships between the objects, to record them in the model. The general procedure is aimed at catching objects and their properties, then catching relationships, one at a time, in isolation from the totality of the scene. This "hunt" continues until the designer completes the description (e.g., class diagram). In the next step, the designer goes astray, should he/she then model activities separately; that is, sequencing, states, blocking, interaction, etc. (e.g., an activity diagram).

The whole modelling process is a search for a grip of the model of the whole system. How can we model the actual process to obtain a better grip?

In object orientation, the designer just dives into a sea of categorization, classes, attributes, and activities, hoping to find a grip on the whole that is constructed from objects tied by relationships, behavior, and structures. The result is not only far from a maximally gripped depiction of the system but, instead, is closer to Tom Hanks's raft design in *Cast Away*, which, admittedly, achieved its goal of reaching civilization. Still, the raft is far different, say, from the maximally gripped design of Polynesian boats used to make voyages across thousands of miles of Pacific Ocean.

According to Merleau-Ponty, when objectifying the world, one tends to view things but fails to see the landscape of Being [68]. The TM approach seems to provide a way to achieve maximum sharpness of the modeler's "perception" of the modeled system and its activities, and it thereby provides a general setting in which the system can exist with the world.

To create a minimal grip, we start with exploration through flows, with the premise that more detail can be accessed from what is found. One thing leads to another (including triggering), which guarantees contiguity in terms of time/place and cause/effect. As Merleau-Ponty probably would have put it, modeling is experienced as a steady flow of things in response to the actual system.

For example, in a business process, if we "catch" an order, then we are blindly led by the flow of an *order*, from a customer to a supplier, which triggers the flow of an *invoic*e, which leads to a flow of *payment*, which in turn leads to the flow of the *ordered product* to the customer. Perceiving things (and their flows and triggers) forms a train of fabric of the developed model, susceptible to furthering a firmer grip of details.

The same model of the system can be enriched, through further refinements, to incorporate more detailed situations and activities into reality, which thus approaches a maximum grip on the modeled part of the world.

## VI. Conclusion

This paper follows from previous research about TM modeling and has enhanced the model with additional concepts such as grand model, maximum grip, and the emphasis that the notion of a thing is a machine and a machine is a thing. Additionally, we contrasted design features of TM with the OO approach. This was not introduced to imply that one way is better than the other. According to GARE [67],

> While Heidegger himself might have concluded that he had no basis for judging the superiority of one way of revealing the world over another, when seen as part of the broader tradition of process thought and the struggle to overcome the mechanistic world view, Heidegger's work can be seen as a contribution to this struggle for a more adequate comprehension of nature and our place within it.

Such ideas can be evaluated according to how they overcome the blind spots and aporias of rival ways of understanding the world. Advancing process thought, Heidegger has also advanced our understanding of what it is to understand the world

In conclusion, this further exploration and enhancement of the TM approach seems to confirm its potential value.